\documentclass[11pt]{article}
\usepackage{amsmath}
\usepackage{pstricks}
\usepackage{slashed}
\usepackage{cite}
\usepackage[vcentermath]{youngtab}
\usepackage{bbm}
\usepackage{amssymb,amsfonts}
\usepackage{relsize}

\newcommand{\re}[1] {(\ref{#1})}
\newcommand{\pb}[2]{\{#1,#2\}}
\newcommand{\pbb}[2]{\{#1,#2\}_B}
\newcommand{\pbf}[2]{\{#1,#2\}_F}

\newcommand{\org}[1]{{\color{orange}{#1}}}

\newcommand{\nn}{\nonumber}

\def\Dsl{\slashed{D}}

\def\half{\frac{1}{2}}
\def\bsh{\backslash}

\newfont{\bbbold}{msbm10 scaled \magstep1}

\def\bbC{\mbox{\bbbold C}}

\def\bbN{\mbox{\bbbold N}}

\def\cA{{\cal A}}
\def\cB{{\cal B}}

\def\cD{{\cal D}}

\def\cL{{\cal L}}

\def\cN{{\cal N}}
\def\cO{{\cal O}}

\newfont{\goth}{eufm10 scaled \magstep1}

\def\gf{\mbox{\goth f}}
\def\gg{\mbox{\goth g}}

\def\gl{\mbox{\goth l}}

\def\go{\mbox{\goth o}}
\def\gp{\mbox{\goth p}}

\def\gs{\mbox{\goth s}}

\def\gu{\mbox{\goth u}}

\def\a{\alpha}\def\adt{\dot \alpha}
\def\b{\beta}\def\bdt{\dot \beta}
\def\c{\gamma}\def\C{\Gamma}
\def\d{\delta}
\def\e{\epsilon}\def\ve{\varepsilon}

\def\h{\eta}

\def\l{\lambda}\def\L{\Lambda}

\def\P{\Pi}

\def\th{\theta}

\def\be{\begin{equation}}\def\ee{\end{equation}}
\def\bea{\begin{eqnarray}}\def\eea{\end{eqnarray}}
\def\barr{\begin{array}}\def\earr{\end{array}}

\def\o{\omega}
\def\del{\partial}

\def\una{\underline a}
\def\unb{\underline b}

\def\xz{\times}

\def\nab{\nabla}

\let\la=\label

\def\nn{\nonumber}
\def\bd{\begin{document}}
\def\ed{\end{document}}
\def\ba{\begin{array}}
\def\ea{\end{array}}
\def\bea{\begin{eqnarray}}
\def\eea{\end{eqnarray}}
\def\ft#1#2{\tfrac{#1}{#2}}
\def\fft#1#2{\frac{#1}{#2}}
\def\sst#1{{\scriptscriptstyle #1}}
\def\oneone{\rlap 1\mkern4mu{\rm l}}

\newcommand{\eq}[1]{(\ref{#1})}
\newcommand{\w}[1]{\\[0.#1cm]}
\newcommand{\ww}{\\[0.0cm]}
\def\eqs#1#2{(\ref{#1}-\ref{#2})}
\def\det{{\rm det\,}}
\def\tr{{\rm tr}}\def\str{{\rm str}}
\def\ad{{\rm ad}}

\newcommand{\hoch}[1]{$\, ^{#1}$}
\newcommand{\imperial}{\it\small Theoretical Physics Group, Imperial College London\\ Prince Consort Road, London SW7 2AZ, UK}
\newcommand{\kings}
{\it\small Department of Mathematics, King's College, University of London\\ Strand, London WC2R 2LS, UK}
\newcommand{\uu}
{\it\small Department of Theoretical Physics, Uppsala, Sweden}
\newcommand{\hip}
{\it\small HIP-Helsinki Institute of Physics, P.O. Box 64 FIN-00014
University of Helsinki, Suomi-Finland}
\newcommand{\stock}
{\it\small Department of Theoretical Physics, Stockholm, Sweden}
\newcommand{\golm}
{\it\small AEI, Max Planck Institut f\"ur Gravitationsphysik\\ Am M\"{u}hlenberg 1, D-14476 Potsdam, Germany}
\makeatletter
\renewcommand\theequation{\thesection.\arabic{equation}}
\@addtoreset{equation}{section} \makeatother

\newcommand{\sa}{/ \hspace{-1.2ex}}
\newcommand{\saa}{/ \hspace{-1.4ex}}
\newcommand{\saaa}{\, / \hspace{-1.6ex}}
\newcommand{\Scal}[1]{\Bigl ({#1} \Bigr )}
\newcommand{\scal}[1]{\bigl ({#1} \bigr )}

\newcommand{\CR}{\nonumber \\*}

\newcommand{\trace}{\hbox {tr}~}
\newcommand{\traceS}{\hbox {tr}_{\scriptscriptstyle \mathfrak{S}}~}

\DeclareMathAlphabet{\mathpzc}{OT1}{pzc}{m}{it}
\def\BRST{\,\mathpzc{s}\,}
\def\aBRST{{\scriptstyle (\mathpzc{s})}}
\def\q{{{\scriptscriptstyle (Q)}}}
\def\qs{{\scriptscriptstyle (Q\mathpzc{s})}}
\def\Qsla{{\mathcal{S}_{\q}}}
\def\Slav{{\mathcal{S}_\aBRST}}
\def\epsilonb{{\overline{\epsilon}}}
\def\bulletup{{\scriptstyle \bullet}}

\newcommand{\gra}[2]{{\scriptscriptstyle (#1 , #2 )}}
\newcommand{\ord}[1]{{\scriptscriptstyle (#1)}}

\def\cL{{\cal L}}
\def\cN{\mathcal{N}}
\def\cO{\mathcal{O}}

\def\ie{{\it i.e.}\ }
\def\eg{{\it e.g.}\ }

\newcommand{\sfrac}[2]{{\scriptstyle \frac{#1}{#2}}}
\newcommand{\stfrac}[2]{{\scriptscriptstyle \frac{#1}{#2}}}

 \def\balpha{{\overline{\alpha}}}
 \def\bbeta{{\overline{\beta}}}
 \def\bgamma{{\overline{\gamma}}}
 \def\bdelta{{\overline{\delta}}}
 \def\bepsilon{{\overline{\epsilon}}}
 \def\bvarepsilon{{\overline{\varepsilon}}}
 \def\bzeta{{\overline{\zeta}}}
 \def\bareta{{\overline{\eta}}}
 \def\btheta{{\overline{\theta}}}
 \def\bvartheta{{\overline{\vartheta}}}
 \def\biota{{\overline{\iota}}}
 \def\bkappa{{\overline{\kappa}}}
 \def\blambda{{\overline{\lambda}}}
 \def\bmu{{\overline{\mu}}}
 \def\bnu{{\overline{\nu}}}
 \def\bxi{{\overline{\xi}}}
 \def\bpi{{\overline{\pi}}}
 \def\brho{{\overline{\rho}}}
 \def\bvarrho{{\overline{\varrho}}}
 \def\bsigma{{\overline{\sigma}}}
 \def\bvarsigma{{\overline{\varsigma}}}
 \def\btau{{\overline{\tau}}}
 \def\bphi{{\overline{\phi}}}
 \def\bvarphi{{\overline{\varphi}}}
 \def\bchi{{\overline{\chi}}}
 \def\bpsi{{\overline{\psi}}}
 \def\bomega{{\overline{\omega}}}

\def\thalf{{\textrm{\tiny\textonehalf}}}
\def\tquarter{{\textrm{\tiny\textonequarter}}}
\def\Ko{{\scriptscriptstyle K}}
\def\tKo{\scriptscriptstyle k }
\def\corr{$\clubsuit$}

\def\ytzero{{}^{\!\raisebox{6pt}{$\mathsmaller{0}$}}}

\newcommand{\auth}{\large P.S.\ Howe${}^{a,}$\footnote{email: paul.howe@kcl.ac.uk} and U. Lindstr\"om${}^{b,c,}$\footnote{email: ulf.lindstrom@physics.uu.se}}

\begin{document}

\renewcommand{\thefootnote}{\fnsymbol{footnote}}

\null
\begin{flushright}
{\small KCL-MTH-02}\\
{\small UUITP-31/18}\\
{\small Imperial-TP-UL-02}\\
\vskip 1.5 cm
\end{flushright}

\begin{center}
{\Large{\bf Some remarks on (super)-conformal Killing-Yano tensors}}
\vspace{.75cm}

\auth
\end{center}
\vspace{.5cm}

\centerline{${}^a${\it \small Department of Mathematics, King's College London}}
\centerline{{\it \small The Strand, London WC2R 2LS, UK}}
\vspace{.5cm}
\centerline{${}^b${\it \small Department of Physics and Astronomy, Theoretical Physics, Uppsala University}}
\centerline{{\it \small SE-751 20 Uppsala, Sweden }}
\vspace{.5cm}
\centerline{${}^c${\it \small Theoretical Physics, Imperial College, London}}
\centerline{{\it \small Prince Consort Road, London SW7 2AZ, UK}}

\vspace{1cm}


\centerline{{\bf Abstract}}
\vskip .5cm
A Killing-Yano tensor is an antisymmetric tensor obeying a first-order differential constraint similar to that obeyed by a Killing vector. In this article we consider generalisations of such objects, focusing on the conformal case. These generalised conformal Killing-Yano tensors are of mixed symmetry type and obey the constraint that the largest irreducible representation of $\go(n)$ contained in the tensor constructed from the first-derivative applied to such an object should vanish. Such tensors appear naturally in the context of spinning particles having $N_0=1$ worldline supersymmetry and in the related problem of higher symmetries of Dirac operators. Generalisations corresponding to extended worldline supersymmetries and to spacetime supersymmetry are discussed.

\vspace{1cm}


\renewcommand{\thefootnote}{\arabic{footnote}}
\setcounter{footnote}{0}

\pagebreak
\tableofcontents
\setcounter{page}{1}


\section{Introduction}

Killing-Yano tensors  \cite{Yano,Sommers} and conformal Killing-Yano tensors \cite{Tac, Krtous:2006qy} are antisymmetric tensors obeying constraints similar to those obeyed by (conformal) Killing tensors. We shall refer to them as (conformal) Killing-Yano forms, and use the term (conformal) Killing-Yano tensors to include these and also mixed-symmetry tensors which are related to them. In flat $n$-dimensional Euclidean space or in spacetime, the constraint on a Killing-Yano $p$-form (KY form) $A$ is given by
\be
\del_{(a} A_{b)c_1\ldots c_{p-1}}=0
\la{1.1}
\ee
whereas that for a conformal Killing-Yano (CKY) form is
\be
\del_{(a} A_{b)c_1\ldots c_{p-1}}=\frac{1}{n-p+1}\left(\h_{ab}\del^d A_{d c_1\ldots c_{p-1} }-
(p-1)\h_{(a[c_1}\del^d A_{|d|b) c_2\ldots c_{p-1]} }\right)\ ,
\la{1.2}
\ee
where $x^a$ are standard flat coordinates and $\h_{ab}$ is the flat metric. Alternatively,
\be
\del_a A_{b_1\ldots b_p}=\del_{[a} A_{b_1\ldots b_p]}+ \frac{p}{n-p+1}\h_{a[b_1}\del^d A_{|d|b_2\ldots b_p]}\ .
\la{1.2.1}
\ee
 It is straightforward to show that the dual of a CKY $p$-form is a CKY $(n-p)$-form, i.e. satisfies the above equation with  $p$ replaced by $n-p$.\black

The generalisation of these equations to the curved case is obtained by replacing the partial derivatives by the standard metric, torsion-free covariant derivative, $\nab$. 

In the curved case KY forms have found applications in general relativity \cite{Carter,Penrose}, to G-structures \cite{Papadopoulos:2007gf,Papadopoulos:2011cz} and string theory backgrounds  {\cite{Chervonyi:2015ima}}, to classical mechanics \cite{Cariglia:2014dfa} and to symmetries of the Dirac operator {\cite{Cariglia:2012ci}}. A comprehensive survey of these topics, together with many more references, can be found in  \cite{Santillan:2011sh}.

In this article we shall concentrate on more formal aspects of CKYTs (conformal Killing-Yano tensors), focusing on their conformal properties. We define CKYTs  to be tensors of the type that can be constructed as highest weight representations arising in products of a conformal Killing tensor (CKT)  and one or more CKY forms, and in the next section we discuss these in some generality in flat spacetime. One motivation for this definition is that such  tensors appear naturally in the context of invariants of spinning particles which have $N_0=1$ supersymmetry on the world-line \cite{vanHolten:1992bu,Gibbons:1993ap}.  This is discussed in section 3, along with an algebra derived from taking the Poisson brackets of two invariant functions determined by CKYTs. We also briefly discuss particles with $N_0=2$ world line supersymmetry and show that higher-rank CKYTs of the type exemplified in \eq{2.5} below arise naturally in this case. In the $N_0=1$ case the worldline supersymmetry gives rise to the Dirac operator $\Dsl$ when quantised, and in section 3 we  also briefly discuss, in five-dimensional spacetime, how higher symmetries of $\Dsl$, in the sense of Eastwood \cite{Eastwood:2002su},  naturally involve CKYTs. The quantisation of an $N_0=2$ particle model gives rise to generalised Maxwell equations, and the CKYTs yield higher symmetries of these equations \cite{Kress}. 
We then go on to investigate extensions of such tensors in superspace. We give definitions of superconformal KYTs (SCKYTs) and look at their properties and examples in spacetime dimensions $4,5,6$.\footnote{We recall that superconformal algebras exist only in these spacetime dimensions; for all values of $N$ in $D=4,6$ and for $N=1$ only in $D=5$ {\cite{Nahm:1977tg,Scheunert:1976uf,Scheunert:1976ug,Kac:1977qb,Kac:1977em,Sohnius:1976pa}}.} In particular, in $D=4$ we relate them to objects briefly discussed in \cite{Howe:2015bdd}, and mention a further generalisation, to SCKYT-spinors. We also discuss SCKYTs in analytic superspaces.  
\section{Conformal Killing-Yano Tensors}

A conformal Killing-Yano $p$-form obeys the constraint given above in \eq{1.2}. In flat $n$-dimensional Euclidean space, in terms of representations of $\go(n)$ this means that the largest representation in the product of a $p$-form multiplied by the derivative one-form must vanish. Moreover, it is not difficult to see that the expansion of the CKY form $A_p$ in powers  of $x$ terminates at $x^2$. 
The components of $A_p$ at $x^0, x^1$ and $x^2$ are given by (constant) antisymmetric tensors of of rank $p, (p-1)+(p+1)$, and $p$ respectively.
These are representations of $\go(n)$ and fit together into the $(p+1)$-form representation of the conformal algebra $\go(1,n+1)$. 

Now suppose one takes a product of a conformal Killing vector $C_a$ and a CKY 2-form $B_{ab}$ and then projects onto the highest weight representation to get
\be
A_{a,bc}:=B_{a(b} C_{c)}+\frac{1}{(n-1)}\left(\h_{a(b} (B\cdot C)_{c)}-\h_{bc} (B\cdot C)_a\right)\ ,
\la{2.1}
\ee
where $(B\cdot C)_a=B_{ab} C^b$. We shall call this object a conformal Killing-Yano tensor (CKYT) of type $(1,2)$. It obeys the constraint that, when one applies a derivative, the highest weight representation in the resultant product of $\go(n)$ representations vanishes. In the following discussion we shall make use of Young tableaux, but with the convention that all the $\go(n)$ or $\go(1,n+1)$ tableaux below refer to tensors that are trace-free.\footnote{This does not necessarily mean that they correspond to irreducible representations because there can be cases where self-duality constraints are also possible; explicit examples of this will be given in the section on $D=6$ superconformal KYTs.}

This construction admits a straightforward generalisation to CKYTs $A_{p,q}$ of type $(p,q)$. Such a tensor has the $\go(n)$ Young tableau
\be
A_{p,q}\sim\ \ \ \ \ {\overbrace{\young(\hfil\hfil\hfil\hfil\hfil,\hfil,\hfil,\hfil)}^{q}}
\la{2.2}
\ee
with $(p+1)$ boxes in the first column.  The differential constraint satisfied by such a CKYT is that, when a derivative is applied to $A_{p,q}$, the traceless tensor corresponding to the Young tableau with one extra box on the first row has to vanish, \ie
\be
\del A_{p,q}\ni \overbrace{\young(\hfil\hfil\hfil\hfil\hfil,\hfil,\hfil,\hfil)}^{q+1}\ = 0\ .
\la{2.2.1}
\ee

These tensors are not new; they appear naturally in the context of the spinning particle \cite{vanHolten:1992bu,Gibbons:1993ap}, as we shall discuss shortly.

It is clear from the diagram \eq{2.2} that we could equally well represent this tableau as a tensor $A'_{p+1,q-1}$, totally antisymmetric on $(p+1)$ indices and symmetric on $(q-1)$ indices such that antisymmetrisation over one further index gives zero. We can define it to be 
\be
A'_{a_1\ldots a_{p+1},b_2\ldots b_q}=(p+1) A_{[a_1\ldots a_p,a_{p+1}]b_2\ldots b_q}\ .
\la{2.2.2}
\ee
Conversely, we have
\be
A_{a_1\ldots a_{p},b_1\ldots b_q}=\frac{q}{p+q} A'_{a_1\ldots a_p (b_1,b_2\ldots b_q)}\ .
\la{2.2.3}
\ee
The $A$-representation turns out to be natural from the point of view of invariants as we shall see in the next section, but the $A'$ version is useful in the context of  duality. Taking the dual of $A'_{p+1,q-1}$ on its antisymmetric indices we get a dual tensor $*A'_{n-p-1,q-1}$, from which we can construct a dual version of $A_{p,q}$ using \eq{2.2.3}. Clearly this tensor will have $n-p-2$ manifestly antisymmetric indices; we shall denote it by $*A_{n-p-2,q}$. Explicitly,
\be
*A_{a_1\ldots a_{n-p-2},b_1\ldots b_q}=\frac{q}{(p+q)}*A'_{a_1\ldots a_{n-p-2}(b_1,b_2\ldots b_q)}\ ,
\la{2.2.4}
\ee
where
\be
*A'_{a_1\ldots a_{n-p-1},b_2\ldots b_q}=\frac{1}{(p+1)!}\ve_{a_{1}\ldots a_{n-p-1}}{}^{c_{1}\ldots c_{p+1}} A'_{c_1\ldots c_{p+1},b_2\ldots b_q}\ .
\la{2.2.5}
\ee
We can therefore write
\be
*A_{a_1\ldots a_{n-p-2},b_1\ldots b_q}=\frac{q}{(p+q)p!}\ve_{a_{1}\ldots a_{n-p-2(b_1}}{}^{c_{1}\ldots c_{p+1}}A_{|c_1\ldots c_p,c_{p+1}|b_2\ldots b_q)}\ .
\la{2.2.6}
\ee

\black

Clearly one can obtain more complicated CKYTs by taking the highest weight in the product of two or more CKY forms and a single CKT (Cartan product). This would lead to tensors $A_{p_1,p_2,\ldots ,q}$ generalising \eq{2.2} with $p_1\geq p_2\geq \ldots$ being the number of boxes minus one in the columns starting from the left. For example, 
\be
A_{p_1,p_2,q}\sim\ \ \ \ \ \overbrace{\young(\hfil\hfil\hfil\hfil\hfil,\hfil\hfil,\hfil\hfil,\hfil,\hfil)}^{q}
\la{2.5}
\ee
where there are $p_1+1$ boxes in the first column and $p_2+1$ in the second. The differential constraint satisfied by this tensor is
\be
\del A_{p_1,p_2,q}\ni \  \overbrace{\young(\hfil\hfil\hfil\hfil\hfil\hfil,\hfil\hfil,\hfil\hfil,\hfil,\hfil)}^{q+1}
= 0\ .
\la{2.5.1}
\ee

The $A_{p,q}$ tensors, like the CKY forms, are representations of the conformal algebra which can be obtained as the highest weight in the product of a CKY $p$-form and and a $(q-1)$th rank CKT. The corresponding Young tableau in $\go(1,n+1)$ is 
\be
A_{p,q}\sim\ \ \ \ \ \overbrace{\young(\hfil\hfil\hfil\hfil\hfil\hfil,\hfil\hfil\hfil\hfil\hfil\hfil,\hfil,\hfil)}^{q}\ ,
\la{2.6}
\ee
where now there are $(p+2)$ boxes in the first column. For $A_{p_1,p_2,q}$ the representation of the conformal algebra is given by
\be
A_{p_1,p_2,q}\sim\ \ \ \ \ \overbrace{\young(\hfil\hfil\hfil\hfil\hfil\hfil,\hfil\hfil\hfil\hfil\hfil\hfil,\hfil\hfil,\hfil\hfil,\hfil,\hfil)}^{q}\ ,
\la{2.7}
\ee
where there are $(p_1+2)$ boxes in the first column and $(p_2+2)$ in the second.

An earlier definition of a generalised conformal Killing-Yano tensor was given by J. Kress in \cite{Kress}. He considered objects which combine CKTs and CKYs in a natural way. These are constructed from the $r$-fold symmetric product of $p$-forms, i.e. they are tensors of the type $K_{a_{11}\ldots a_{1p},a_{21}\ldots a_{2p},\dots ,a_{r1}\ldots a_{rp}}$, antisymmetric on each set of $p$ indices and symmetric under the interchange of any two such sets. In addition, all traces are taken to vanish. These tensors are taken to satisfy certain first-order differential constraints which are given explicitly in \cite{Kress}. This definition is designed to reduce to those for CKTs for $p=1$ and to those for CKYs for $r=1$. These tensors are not irreducible in general but can of course be decomposed into irreducible components, and the latter will be tensors of the sort discussed above.
\black

\section{Spinning particles}

A particle with local worldline supersymmetry provides a classical model for a spinning particle that when quantised gives rise to the Dirac equation  \cite{Berezin:1976eg,Casalbuoni:1975hx,Barducci:1976qu,Brink:1976sz,Brink:1976uf}.
In this section we consider particles with $N_0$ world-line supersymmetries and show that supersymmetric invariants are determined by CKYTs of the type discussed above. We focus on the $N_0=1,2$ cases because models with $N_0\geq 3$ do not admit general background spacetimes, only flat ones, although in the current paper we focus on the flat case for simplicity of presentation.

\subsection{Basics}

The Lagrangian for a spinning particle in flat $n$-dimensional spacetime with $N_0$ local worldline supersymmetries is given by {\cite{Gershun:1979fb,Howe:1988ft}}
\be
L=\dot{x}\cdot p+\frac{i}{2}\l^i\cdot \dot\l_i -\half e p^2 -i\psi^i \l_i\cdot p-\frac{i}{2} f^{ij} \l_i\cdot\l_j\ ,
\la{3.1}
\ee
where $(x^a,p_a,\l^a_i)$ denote the particle's position, momentum and fermionic coordinates, with $a$ a spacetime vector index and $i=1,\ldots N_0$ a vector index for the internal symmetry group $O(N_0)$. The additional variables $(e,\psi_i,f_{ij})$ are the fields of the worldline ``supergravity'' multiplet consisting of the einbein, $e$, $N_0$ einbini, $\psi_i$, and $\half N_0(N_0-1)$ einbosons, $f_{ij}$, gauge fields for the local $O(N_0)$ symmetry. These fields act as Lagrange multipliers imposing the constraints that the Hamiltonian, the supercharges and the $O(N_0)$ currents should vanish. Explicitly, these are respectively
\begin{align}
H&= \half p^2 \nn\w1
Q_i&= \l_i\cdot p\nn\w1
M_{ij}&=i \l_i \cdot \l_j \ .
\label{3.2}
\end{align}
The Lagrangian \eq{3.1} is already in first-order form and comes with a symplectic form $\o$ given by
\be
\o=dx^a\wedge dp_a -\frac{i}{2} d\l^a\wedge d\l_a\ .
\la{3.3}
\ee
The associated Poisson brackets are
\be
\{F,G\}=\frac{\del F}{\del x^a}\frac{\del G}{\del p_a}-\frac{\del F}{\del p_a}\frac{\del G}{\del x^a}+i(-1)^f \frac{\del F}{\del\l^a}\frac{\del G}{\del\l_a}=(-1)^{fg+1}\{G,F\}\ ,
\la{3.4}
\ee
where the indicators $f,g$ are $0,1$ when $F,G$ are even (odd) respectively.  The basic non-zero Poisson brackets are 
\be
\{x^a,p_b\}=\d^a_b\ ,\qquad  \{\l^a_i,\l^b_j\}=-i\d_{ij}\h^{ab}\ ,
\la{3.5}
\ee
while the non-vanishing Poisson brackets for the constraints are 
\begin{align}
\{Q_i,Q_j\}&=-2i\d_{ij} H\nn\w1
\{M_{ij},Q_k\}&=-2\d_{k[i} Q_{j]} \nn\w1
\{M_{ij},M^{kl}\}&=-4i\d_{[i}{}^{[k} M_{j]}{}^{l]} \ .
\label{3.6}
\end{align}
The Poisson brackets of the supersymmetry generators with the particle variables are
\begin{align}
\{Q_i,x^a\}&=-\l^a_i\nn\w1
\{Q_i,\l^a_j\}&=-i\d_{ij}p^a\nn\w1
\{Q_i,p_a\}&=0\ .
\label{3.7}
\end{align}

The action defined by the above Lagrangian \eq{3.1} is invariant under the local worldline symmetries that we have detailed and is also invariant under spacetime conformal transformations. However, it can only be extended to general spacetime backgrounds for $N_0\leq 2$ {\cite{Howe:1989vn}}.\footnote{Backgrounds with constant curvature are allowed, however,  see \cite{Kuzenko:1995mg}.} For this reason we shall focus on these two cases. 

\subsection{$N_0=1$ worldline supersymmetry}

Functions of the form
\be
F=F(x,\l)^{b_1\ldots b_q}p_{b_1}\ldots p_{b_q}
\la{3.8}
\ee
can be expanded in the odd variables to give a sum of terms of the form
\be
\l^{a_1\ldots a_p} A_{a_1\ldots a_p,b_1\ldots b_q}\, p^{b_1 \ldots b_q}\ ,
\la{3.9}
\ee
where the multi-index $\l$ and $p$ denote $p$-fold and $q$-fold products of the odd coordinates and the even momenta respectively.
The coefficient functions are totally antisymmetric on their $a$ indices and totally symmetric on the $b$ indices.  A world-line super-invariant is a function $F$ of the phase-space variables which is weakly annihilated by $Q$, $\{Q,F\}\approx 0$. Since $\{Q,Q\}\sim H\sim \frac{d}{d t}$ such a function will automatically be a constant of the motion modulo the constraints.  Invariants of spinning particles were considered previously  in \cite{vanHolten:1992bu,Gibbons:1993ap}. These authors discussed spinning particles in general gravitational backgrounds but with rigid supersymmetry, rather than local. This means that the particles they consider are not necessarily massless and hence the associated invariant tensors need not be conformal.  Here, we shall show, in the flat case, how the invariants in the massless case  are related to CKYTs of the type $A_{p,q}$ discussed in the previous section. 

The claim is that, given a function whose leading (i.e. lowest order in $\l$) term involves $\l^p p^q$, then there is an invariant $F$ whose leading term has  a CKYT $A_{p,q}$ as its coefficient and which requires only one correction at order $\l^{p+2} p^{q-1}$.  $A_{p,q}$ can be taken to be completely traceless because any trace terms will involve the constraints. The complete expression for $F$ is
\be
F=\l^{a_1\ldots a_p} A_{a_1\ldots a_p,b_1\ldots b_q}p^{b_1\ldots b_q} +i\frac{(-1)^{(p+1)} q}{(1+p+q)}\l^{a_1\ldots a_{p+2}}\del_{a_1} A_{a_2\ldots a_{p+1},a_{p+2}b_2\ldots b_q} p^{b_2\ldots b_q}\ .
\la{3.10}
\ee
When $Q$ is applied to $F$ one finds terms of the form $\l^{p-1} p^{q+1},\ \l^{p+1} p^q,\ \l^{p+3} p^{q-1}$ which must all be zero if $\{Q,F\}\approx 0$. The first of these only involves the leading term and implies that
\be\label{cons1}
A_{a_1\ldots a_{p-1}(a_p,b_1\ldots b_{q})}=0\ .
\la{3.11}
\ee
\black
This means that $A_{p,q}$ is indeed in the representation \eq{2.2}. The $\l^{p+1}p^q$ term is 
\be
\l^{a_1\ldots a_{p+1}}p^{b_1\ldots b_q}\left(\del_{a_1}A_{a_2\ldots a_{p+1},b_1\ldots b_q}-\frac{q(p+2)}{(1+p+q)} \del_{[a_1}A_{a_2\ldots a_{p+1},b_1]\ldots b_q}\right)\ ,
\la{3.12}
\ee
where the anti-symmetrisation over $(p+2)\ a$-type  indices in the second term comes from the fact that the second term on the right in \eq{3.10} has $(p+2)\ \l$s. Using \eq{3.11} one can then rearrange the terms in \eq{3.12} to find that
\be\label{cons2}
\del_{(a_1}{A_{|a_2\ldots a_{p+1}|,b_1\ldots b_q)}}_{| 0}=0\ ,
\la{3.13}
\ee
where the zero subscript indicates the trace-free part and where the terms inside the vertical bars are excluded from the symmetrisation. This is just the differential constraint \eq{2.2.1} on $A_{p,q}$. Finally, the term with $\l^{p+3}p^{q-1}$ is clearly zero since it involves two derivatives acting on $A$ which are anti-symmetrised because they are contracted with $\l$ indices.

The above discussion is in fact quite general for the systems we consider. Clearly, invariants can be divided into two classes according to whether they are Grassmann even or odd. For the former the fewest number of $\l$s is 0, while for the latter it is 1. Since the powers of momenta do not increase as we repeatedly apply $Q$ to a given term in an invariant, it follows that we can consider invariants whose lowest-possible order terms have a particular power, $q$ say, of momenta. It is easy to see that these will either be $q$th rank CKTs or CKYTs of type $(1,q)$ for even or odd types respectively. If one now examines the conditions for the entire sequence of terms for a given leading term to be invariant one finds that at each step there are two representations that can arise. One of these is determined in terms of the derivative of the previous term, while the other is a new term in  an irreducible representation of the type \eq{2.2}. In addition, the derivative of the previous term must satisfy the differential constraint \eq{2.2.1}. This tensor will then not contribute to the succeeding term because the two derivatives acting on it are antisymmetrised, as in the example above, while the new undetermined tensor will be the leading $(1,q-1)$ or $(2,q-1)$ term in an invariant of the type \eq{3.10}. In short, a general invariant can be decomposed into a sequence of invariants of the type that we have discussed above, at least in the free case.

As we have seen earlier the number of independent types of $A_{p,q}$s that exist is reduced by duality. For $n$ even it is $\frac{n}{2}$ while for $n$ odd it is $\frac{n-1}{2}$. For example, for $n=5$ we have the following independent types: $A_{0,q}$, which are simply CKTs, and $A_{1,q}$. $A_{2,q}$ can be obtained by duality from $A_{1,q}$ while $A_{3,q}$ can be  obtained from $A_{0,q}$ and is therefore  another representation of a CKT. Finally, $A_{4,q}$ is formally dual to $A_{-1,q}$ which can be interpreted as a CKT of rank $q-1$. For the case of $n$ even there is also the possibility of self-dual $A_{p,q}$ tensors for $p=\frac{n-2}{2}$. In the classical theory, different powers of $\l$ are deemed to be independent so that the same $A_{p,q}$ tensors can arise for different powers of $\l$ according to the rules derived from duality. However, the situation is not the same in the quantum theory as we shall discuss below.
\black

{We shall abbreviate the above expression \eq{3.10} for the invariant $F$ by
\be
F=\l^p A_{p,q}p^q+\a(p,q)\l^{p+2} dA_{p+2,q-1} p^{q-1}:=A+dA
\la{3.14}
\ee
where 
\be
\a(p,q):= i\frac{(-1)^{(p+1)} q}{(1+p+q)}\ ,
\la{3.15}
\ee
\be
\left(dA_{p+2,q-1}\right)_{a_1\ldots a_{p+2},b_1\ldots b_{q-1}}:=\del_{[a_1}A_{a_2\ldots a_{p+1},a_{p+2}]b_1\ldots b_{q-1}}~,\ 
\la{3.16}
\ee
and, {by abuse of} notation, $dA$ is used to remind us that, for $N_0=1$, the correction term to a correction term is zero, $d^2A=0$.} (Alternatively the tensor in the second term on the right-hand side of \eq{3.16} could be written $(dA')_{p+2,q-1}$ where $A'_{p+1,q-1}$ is given in \eq{2.2.2} and the $d$ acts on the form indices only.) 

Consider the Poisson bracket of two invariant functions $F$ and $G$ with leading terms $A$ and $B$, respectively. 
It follows from the Jacobi identity that this will produce new invariant functions. The Poisson bracket
\bea\label{pbfg}
\pb{F}{G}=\pb{A+dA}{B+dB}
\eea
 will produce terms of type  $(p+p'-2,q+q')$, $(p+p', q+q'-1)$,  $(p+p'+2, q+q'-2)$  and  $(p+p'+4, q+q'-3)$.  These will result in four new invariants $H^{(1)},..., H^{(4)}$, with leading terms $C^{(1)},..., C^{(4)}$ of the aforementioned type. 
 
 There is a unique $(p+p'-2,q+q')$ term and corresponding  $H^{(1)}$\footnote{The subscripts $B$ and $F$ below refer to the bosonic ($x,p$) and fermionic ($\l$) parts of the Poisson bracket respectively.}
 \bea\nn
 &&C^{(1)}=\pbf{\l A p}{\l Bp}\\[1mm]
 &&H^{(1)}=C^{(1)}+dC^{(1)}~.
 \eea
 The correction term $dC^{(1)}$  is of type $(p+p', q+q'-1)$, and we find the next leading term by subtracting it from the other terms of this type generated in \re{pbfg}:
 \bea\nn
&& C^{(2)}=-dC^{(1)}+\pbb{\l Ap}{\l Bp}+\pbf{\l Ap}{\alpha \l dBp}+\pbf{\alpha\l dAp}{ \l Bp}\\[1mm]
&& H^{(2)}=C^{(2)}+dC^{(2)}~.
 \eea
 The correction term $dC^{(2)}$ is now of   type $(p+p'+2, q+q'-2)$  and we subtract it from the other terms of this type
 \bea\nn
 &&C^{(3)}=-dC^{(2)}+\pbb{\l Ap}{\alpha\l dB p}+\pbb{\alpha\l dAp}{ \l Bp}+\pbf{\alpha\l dAp}{ \alpha\l dBp}\\[1mm]
 && H^{(3)}=C^{(3)}+dC^{(3)}~.
 \eea
 The correction term $dC^{(3)}$ is now of  type $(p+p'+4, q+q'-3)$  and we subtract it from the one remaining term from \re{pbfg}, which is of this type
 \bea\nn
 &&C^{(4)}=-dC^{(3)}+\pbb{\alpha\l dAp}{\alpha\l dB p}\\[1mm]
 && H^{(4)}=C^{(4)}+dC^{(4)}~.
 \eea
Consistency requires that $H^{(4)}$ does not require a correction term, i.e., that $$dC^{(4)}=d\pbb{\alpha\l dAp}{\alpha\l dB p}=0$$. This is indeed the case, due to products of partial derivatives being anti symmetrised by $\l$s.

One may work out the explicit tensor expressions  in for the various $C$s and $H$s. E.g.,
\bea
\pbf{\l A p}{\l Bp}=i(-)^p (pp') \l^{p+p'-2}A_{cp-1,q}\delta^{cd}B_{dp'-1,q'}p^{q+q'}
\eea
where the numerical powers brought down from $\l$s are  put in parentheses so as not to be confused with momenta. Although the resulting expressions for some terms may be simplified using the constraints
\re{cons1} and \re{cons2}, the explicit expressions are not terribly illuminating. Invariance of the $H$s follow by construction, but is of course also possible to check explicitly.
We have done so, along with a check of the constraints, in the simple case of $(p,q)=(p',q')=(1,1)$, where only  $H^{(1)}$ and $H^{(2)}$ are nonzero.

Thus the Poisson bracket gives rise to Lie algebra structure on the space of CKYTs of type $(p,q)$.

\subsection{$N_0=2$ worldline supersymmetry}

In the $N_0=2$ case it is convenient to use complex notation, so we set $\xi=\frac{1}{\sqrt{2}}(\l_1+i\l_2)$, and similarly for $Q$. We then have
\begin{align}
\{Q,\xi^a\}&=\{\bar Q,\bar\xi^a\}=0;\qquad \{Q,\bar\xi^a\}= \{\bar Q,\xi^a\}=-ip^a\nn\w1
\{Q,x^a\}&=-\xi^a;\qquad \{\bar Q,x^a\}=-\bar\xi^a\nn\w1
\{Q,Q\}&=\{\bar Q,\bar Q\}=0;\qquad \{Q,\bar Q\}=-2H \ ,
\label{34}
\end{align}
as well as
\be
\{\xi^a,\xi^b\}=\{\bar\xi^a,\bar\xi^b\}=0;\qquad \{\xi^a,\bar\xi^b\}=-i\h^{ab}\ .
\la{}
\ee
Super-invariants $F$ will have leading terms of the form
\be
F^{(0)}=\xi^{a_1\ldots a_m}\bar\xi^{b_1\ldots b_m} B_{a_1\ldots a_m,b_1\ldots b_m,c_1\ldots c_q} p^{c_1\ldots c_q}\ .
\la{3.25}
\ee
Clearly there must be the same number of $\xi$s and $\bar\xi$s in order to maintain $U(1)$ symmetry. As in the $N_0=1$ case we may assume that $B$ is completely traceless because any trace terms will be proportional to constraints. In order to be invariant under $O(2)$ and not just $SO(2)$ one also  requires symmetry under $\l_1\mapsto \l_2,\ \l_2\mapsto-\l_1$, or $\xi \leftrightarrow\bar\xi$. This requires
\be
B_{\una,\unb}=(-1)^m B_{\unb,\una}\ ,
\la{3.26}
\ee
where $\una(\unb)$ denote the sets of $a(b)$ indices and where the $c$ indices have been ignored. If we further require $F$ to be real then $B$ is real and symmetric for $m$ even and imaginary and antisymmetric for $m$ odd.  So the conclusion is that the tensors appearing as leading terms in $N_0=2$ super-invariants have two sets of $m$ antisymmetrised indices as well as $q$ symmetrised indices, and are also traceless. Moreover, supersymmetry at the lowest order (i.e. acting on the fermions) implies that symmetrisation over more than $q$ indices gives zero. The representations contained in $B$ are not in general irreducible but this constraint implies that $B$ can be decomposed into  irreducible tensors of type $A_{p_1,p_2,q}$ of section 2.  We therefore have
\be
B_{m,m,q}=\sum_{r=0}^{r=k} A_{2m-2r,2r,q}\ ;\qquad  m =\begin{cases} 2k& \text{for}\  m \ \text{even}\\ 
2k+1 &\text{for}\ m\  \text{odd}\end{cases}\ ,
\la{3.27}
\ee
where on the left-hand side $B_{m,m,q}$ denotes the function appearing in $F$ before decomposition into irreducibles.

The full classical invariant is obtained by adding $\xi^{m+1}\bar\xi^{m+1} p^{q-1}$ and $\xi^{m+2}\bar\xi^{m+2} p^{q-2}$ terms to \eq{}. Explicitly,
\be
F^{(1)}=\xi^{a_1\ldots a_{m+1}}\bar\xi^{b_1\ldots b_{m+1}} B^{(1)}_{a_1\ldots a_{m+1},b_1\ldots b_{m+1},c_2\ldots c_q} p^{c_2\ldots c_q}\ ,
\ee
where
\begin{align}
B^{(1)}_{a_1\ldots a_{m+1},b_1\ldots b_{m+1},c_2\ldots c_q}&=\left(\sum_{r=0}^{r=k}\a_r \del_{a_1}B_{a_2\ldots a_s b_1\ldots b_r,b_{r+1}\ldots b_{m} a_{s+1}\ldots a_{m+1},b_{m+1}c_2\ldots c_q}\right.\nn\w1
&\left.+\sum_{r+1}^{r=[k]}\b_r \del_{a_1}B_{a_2\ldots a_s b_1\ldots b_r,b_{r+1}\ldots b_{m+1} a_{s+1}\ldots a_m,a_{m+1}c_2\ldots c_q}\right)\nn\w1
&+(-1)^{m+1} \una\leftrightarrow\unb
\end{align}
where $[k]=k$, if $m$ is even,  or $ (k+1)$ if $m$ is odd. On the right-hand-side of this equation the $a$s and $b$s are separately antisymmetrised. The constants $\a_r,\b_r$ are determined by requiring invariance under $Q$. When $Q$ is applied to $F^{(0)}$ there will be a term arising from $Q$ being applied to $B$ which will bring down a factor of $\xi$ together with a spacetime derivative acting on $B$. Moreover, the terms coming from $B^{(1)}$ when $Q$ is applied to a $\bar\xi$, converting it to a $p$, will give terms of the same structure. This term will have to be cancelled by the $\a_0$ term in $B^{(1)}$, while the remaining terms are required by symmetry. It is not difficult to see that there are exactly the right number of constraints to determine the $\a$s and $\b$s, but their precise values are not needed for the current discussion.

In addition, the invariance of $F$ at this level also imposes a differential constraint on $B$:
\be
\del_{(c_0} {B_{|a_1\ldots a_m,b_1\ldots b_m |,c_1\ldots c_q)}}_{| 0}=0\ .
\ee
When $B$ is decomposed into irreducibles, as in \eq{3.27}, this constraint implies the one given in \eq{2.5.1} for each $A_{p_1,p_2,q}$ tensor.

If $Q$ and $\bar Q$ are applied to $F^{(0)}$ two derivatives will be brought down, and these can be associated with the $\xi$ and $\bar\xi$ indices so that they will not give zero in contrast to the $N_0=1$ case. This means that a second-order term will be required to complete the picture for $N_0=2$. It is given by
\be
F^{(2)}=\xi^{a_1\ldots a_{m+2}}\bar\xi^{b_1\ldots b_{m+2}} B^{(2)}_{a_1\ldots a_{m+2},b_1\ldots b_{m+2},c_3\ldots c_q} p^{c_2\ldots c_q}\ ,
\ee
where
\be
B^{(2)}_{a_1\ldots a_{m+2},b_1\ldots b_{m+2},c_3\ldots c_q}=\sum_{r=1}^{r=[k]} \c_r \del_{a_1}\del_{b_1} B_{a_2\ldots a_r b_2\ldots b_s,a_{r+1}\ldots a_{m+1}b_{s+1}\ldots b_{m+1},a_{m+2}b_{m+2}c_3\ldots c_q}
\ee
and where, as before, $[k]=k$, if $m$ is even,  or $ (k+1)$ if $m$ is odd. In this case note that $B^{(2)}$ automatically has the right symmetry properties under the interchange of $\una$ and $\unb$ indices.

As in the $N_0=1$ case the Jacobi identity guarantees that the Poisson bracket of two invariant functions of the above type is again invariant. In the $N_0=2$ case the calculation is in principle straightforward, but is considerably lengthier and we omit the details.

\subsection{Quantisation}
\subsubsection{$N_0=1$}

The canonical quantisation of the Poisson brackets for $\l^a$ for $N_0=1$ implies that in the quantum theory $\l^a\rightsquigarrow \c^a$, so that the wave function can be identified as a Dirac spinor $\Psi$ and the supercharge becomes the Dirac operator, as is well-known. The quantum versions of the invariants determined by the CKYTs will then define the leading terms in higher symmetries of the Dirac equation in a similar way to which CKTs define higher symmetries of the Laplacian (or massless wave equation) in the purely bosonic case. As in that case there is some ambiguity in the lower-order terms that can accompany the leading ones which from this point of view can be seen as operator-ordering problems. Although there are canonical choices in the case of the Laplacian  \cite{Eastwood:2002su} (and super-Laplacians \cite{Howe:2016iqw}) to our knowledge there has not been a similarly complete study carried out in the case of the Dirac operator, although there are some results in the literature for particular cases \cite{Santillan:2011sh} (and references therein). One complication in the Dirac case is that there can be different types of spinor depending on the dimension of spacetime. In the pseudo-classsical formalism decribed above one can introduce further constraints that allow one to impose chirality or Majorana conditions on the wave function (or both) {\cite{Howe:1989vn}}, but there are also symplectic constraints that can be introduced which would require further changes to the classical Lagrangian.

To illustrate the formalism we consider the case of $D=5$ spacetime with metric $-dt^2+ d\underline{x}^2$. Dirac spinors are 4-component complex which implies that the symmetries we wish to consider will involve complex tensors in general. Given two 4-component Dirac spinors one could impose a symplectic-Majorana reality condition which also gives rise to spinors with 8 real components, but this introduces an additional $\gs\gp(1)$ symmetry so that one would have to consider tensors carrying representations of this algebra as well as spacetime indices. We shall therefore stick to the case of a single Dirac spinor for simplicity.

We recall that a differential operator $\cD$ of degree $q$ is a higher-order symmetry of the Dirac operator $\Dsl$ if
\be
\Dsl \cD=\d \Dsl
\la{}
\ee
for some other differential operator $\d$, in other words $\cD\psi$ is a solution of the Dirac equation if $\psi$ is. 

Consider the case of a first-order symmetry. Its most general form would be
\be
\cD=(A^a + \c^b A_b{}^a + \half \c^{bc} A_{bc}{}^a)\del_a + (B+\c^a B_a + \half \c^{ab} B_{ab})\ .
\la{}
\ee
The first observation is that both of the $A$-tensors that accompany $\c$-matrices in the derivative term can be taken to be trace-free because their traces would lead to $\Dsl$ terms. Applying $\Dsl=\c^a\del_a$ in the flat case we get terms with 0,1 or 2 derivatives on the right. The term with two derivatives implies that both of these $A$ tensors are totally antisymmetric, i.e. they are respectively 2- and 3-forms. But now we can replace the 3-form with its dual and dualise the $\c_2$-matrix to get a contribution to $\cD$ of the form
\be
\c^{abc} *A_{ab} \del_c \sim \c^a *A_a{}^b \del_b
\la{}
\ee
after dropping a $\Dsl$ term. So the 3-form $A$-term in $\cD$ can be absorbed into the 2-form term. This implies that there are just two independent tensors in the leading term of $\cD$. It is not difficult to check that $A^a$ is simply a CKV, so we can focus on $A_a{}^b$ which will require associated $B$ terms. The terms in $\Dsl\cD$ with one $\del$ determine $B_a$ and $B_{ab}$ in terms of a derivative acting on $A_{ab}$. This tensor  has 3 irreducible components, a 3-form, a divergence $(\del\cdot A)_a=\del^b A_{ba}$, and a trace-free mixed symmetry tensor. The latter does make a contribution to the one derivative terms, but clearly cannot be absorbed by the $B$s, so that we obtain the constraint that $A_{ab}$ is indeed a CKY 2-form. For the $B$s we find
\be
B_a\sim (\del \cdot A)_a\ ;\qquad B_{ab}\sim i\ve_{abcde} \del^c A^{de}\ ,
\la{}
\ee
while the scalar $B$ is a constant.
Finally, one can check that the terms in $\Dsl\cD$ with no derivatives are satisfied by virtue of the derivative constraint on $A$.

Similar analyses can be applied to $q$th order symmetries. There will be those with $q$th order CKTs and those with leading terms of the form
\be
\cD\sim \c^a A_a{}^{b_1\ldots b_q}\del_{b_1\ldots b_q}\ .
\la{}
\ee
Note that we can take the $q$  derivatives on the right to be traceless because any  trace would give the square of $\Dsl$. In addition, we can take $A$ to be traceless as a trace term would lead to a contribution of the form $\Dsl \del^{(q-1)}$. Applying $\Dsl$ and looking at the terms with $(q+1)$ derivatives we find that the totally symmetric part of $A$ must vanish so that $A$ is algebraically an $A_{1,q}$-tensor. The terms with $q$ derivatives then lead to the differential constraint \eq{2.2.1} on $A$, as well as determining the $(q-1)$th order terms in $\cD$ (which will involve all the independent gamma-matrices). Completing the calculation to determine all the components of $\cD$ is a lengthy procedure which we shall not discuss further here.

\black

\subsubsection{$N_0=2$}
If we set $f_{ij}=\e_{ij} f$ for $N_0=2$, the Lagrangian \eq{} becomes, using complex notation,
\be
L=\dot x\cdot p+i\bar\xi\cdot\dot\xi-i\psi \bar\xi\cdot p-i\bar\psi\xi\cdot p-\half ep^2 +\half f[\xi,\cdot\bar\xi] \ .
\la{}
\ee
Applying the rules of canonical quantisation we find, in particular, that the anti-commutator of $\xi^a$ and $\bar\xi_b$ is $\d^a_b$, so that we can choose a representation such that the wave-function $\Psi=\Psi(x,\xi)$ and $\bar\xi_a\rightsquigarrow\frac{\del}{\del\xi^a}$. Expanding $\Psi$ in powers of $\xi^a$ we get an inhomogeneous differential form. However, we have to impose the constraint imposed by the Lagrange multipler $f$ which becomes, in the quantum theory,\footnote{For a more detailed discussion of this point see \cite{Howe:1989vn}.}
\be
[\xi^a,\bar\xi_a]\Psi=(\xi^a \frac{\del}{\del\xi^a}-\frac{\del}{\del\xi^a} \xi^a)\Psi\implies \xi^a\frac{\del}{\del\xi^a}\Psi=\frac{n}{2}\Psi\ .
\ee
This then implies that the dimension of spacetime must be even and that $\Psi$ is a differential form of degree $\frac{n}{2}$:
\be
\Psi=\frac{1}{m!}\xi^{a_1\ldots a_m} F_{a_1\ldots a_m},\quad {\rm for} \quad m=\frac{n}{2}\ .
\ee
It is straightforward to see that $Q$ becomes  the exterior derivative $d$ acting on $F$, while $\bar Q$ becomes the divergence operator $\d=* d*$. Thus the wave-function  is an $\frac{n}{2}$-form which is closed and co-closed, i.e. an on-shell abelian gauge field-strength form of degree equal to half the spacetime dimension. We shall refer to such equations as generalised Maxwell equations. 

The invariant functions in the $N_0=2$ case therefore become higher symmetries of generalised Maxwell equations in the quantum theory. Moreover, although the gauge fields themselves are abelian these models are in general compatible with curved background spacetimes, one has to examine any given spacetime explicitly to determine if it admits CKYT tensors which give rise to such higher symmetries. Some non-trivial four-dimensional examples of symmetries of this type have been given in \cite{Kress} (and references therein). 
\section{Superconformal Killing-Yano tensors}

\subsection{General discussion}

Flat super Minkowski space has standard (even, odd) coordinates $(x^a,\th^{\a})$, where $a$ runs from $0$ to $(D-1)$, the dimension of spacetime, and $\a$ is a combined spinor-internal symmetry index, running from $1$ to $N$ times the dimension of the basic spinor representations, and where $N$ denotes the number of supersymmetries. The basic derivatives are $(\del_a, D_{\a})$, with the non-trivial commutation relation
\be
[D_{\a}, D_{\b}]=-i(\C^a)_{\a\b}\del_a
\la{4.1}
\ee
where the $\C^a$ matrices are symmetric and are a product of spinor matrices $\c^a$ with an appropriate internal symmetry invariant. The bracket here is graded anti-symmetric which means that it is an anticommutator for two odd objects. When we need to be specific about the internal symmetry group we shall switch notation $\a\rightarrow \a i$, where $i$ is the internal (R)-symmetry index. 

In flat superspace a rank $p$ superconformal Killing-Yano form (SCKY) form is determined by an antisymmetric even tensor $A_{a_1\ldots a_p}$ satisfying the constraint
\be
D_{\a} A_{a_1\ldots a_p}=(\C_{a_1\ldots a_p} \L)_{\a}\ ,
\la{4.2}
\ee
where $\C$-matrices with multiple even indices are anti-symmetrised products of (in this case) $p$ $\C$-matrices, and $\L$ is a spinor. This should be compared with the constraint for a $q$th rank superconformal Killing tensor (SCKT) $K^{b_1\ldots b_q}$:
\be
D_{\a} K^{b_1\ldots b_q}=(\C^{(b_1}\L^{b_2\ldots b_q)})_{\a}\ ,
\la{4.3}
\ee
where tracelessness on the right-hand-side is assured by taking  the symmetric tensor-spinor $\L$ to be $\C$-traceless. We can check that the definition \eq{4.2} makes sense by applying a second $D$-derivative, using the relation \eq{4.1} and tracing over the spinorial indices. This reproduces the even CKY form constraint on the leading  component of $A$.\footnote{This is not quite true for $D=6$ as we shall discuss in 5.3}

This construction can be be generalised to super $A_{p,q}$ tensors. The basic constraint is taken to be
\be
D A_{a_1\ldots a_p,b_1\ldots b_q}=[\C_{(b_1} \L_{|a_1\ldots a_p|,b_2\ldots b_q)}]\ ,\ q\geq 2\ ,
\la{4.4}
\ee
where we have surpressed the spinor indices on $D$ and $\L$, and where both $A$ and $\L$ are symmetric on the $b$-indices, antisymmetric on the $a$-indices and completely traceless with respect to the even metric. (The bars indicate indices excluded from the symmetrisation.) The square brackets on the right indicate that the traces are to be removed. In this construction there are gauge invariances on the right, so that the number of components in $\L$ is not immediately obvious. To see this, let $S^{p,q}$ denote the space of $A_{p,q}$-valued spinors, i.e. objects that have the same symmetry properties as $A_{p,q}$ but which also carry an extra spinor index. We can define an algebraic differential $\d:S^{p,q}\rightarrow S^{p+1,q}$ by
\be
\d  \L_{a_1\ldots a_p,b_1\ldots b_q}= [\C_{(b_1} \L_{|a_1\ldots a_p|,b_2\ldots b_{(q+1))}}]
\la{4.5}
\ee
Clearly $\d^2=0$. Moreover, there is no cohomology. This enables us to compute the dimension of $\L$ systematically.

For $q=2$, we have $\L_{p,1}$ on the the right-hand side of \eq{4.3} which in our conventions means that $\L$ is a $(p+1)$-form valued spinor. So the $\d$ gauge-invariance does not apply in this case. However, it is easy to see that the right-hand side of \eq{4.3} is invariant if $\L_{p,1}$ is changed by $\C_{p+1}\l$, for some spinor $\l$. Given this we can state a formula for the dimension of the representation appearing on the right of \eq{4.3} modulo gauge invariances. It is:
\be
Dim\ [\L_{p,q-1}]=\left(\sum_{r=1}^{q-2} (-1)^{r+1} d_{p.q-r}\right) \xz d_s\ ,
\la{4.6}
\ee
where $d_{p,q}$ is the dimension of the representation of $A_{p,q}$ and $d_s$ is the dimension of the spinor representation carried by the $\L$s. In the final term, $r=q-1$ we have $d_{p,1}$ which has to be set equal to 1, corresponding to the invariance in {\eq{4.5}}.
\subsection{$D=4$}

In $D=4$ it is convenient to represent CKTs in two-component spinor notation. Thus, an $n$th rank CKT $K^{\a_1\ldots \a_n,\adt_1\ldots \adt_n}$ has $n$ undotted and $n$ dotted spinor indices and is  totally symmetric on both sets. The CKT constraint is
\be
\del_{(\a(\adt} K_{\b_1\ldots\b_n),\bdt_1\ldots\bdt_n)}=0\ .
\la{4.6}
\ee
Similarly, an $n$th rank SCKT is an object defined on superspace with the same index structure but obeying the constraint
\be
D_{(\a i} K_{\b_1\ldots\b_n),\bdt_1\ldots\bdt_n}=0
\la{4.7}
\ee
together with its complex conjugate involving $\bar D_{\adt}^i$ and symmetrisation over the dotted indices. The R-symmetry indices refer to the $N$ and $\bar N$ representations of $SU(N)$.  

As noted in \cite{Howe:2015bdd}, one can also consider objects $K^{m,n}, m\neq n$, which obey similar constraints, either in spacetime or in superspace. For $m+n$ even these will be tensorial, while for $m+n$ odd we get spinorial objects. The latter might be called conformal or superconformal KYT spinors. 

In $D=4$ it only makes sense to consider CKY 2-forms, since 3-forms are dual to 1-forms and the latter are equivalent to CKVs. We thus have $K^{(2,0)}$ together with its conjugate $\bar K^{0,2}$. In the even case the constraint is 
\be
\del_{(\a \adt} K_{\b\c)}=0
\la{4.8}
\ee
which, together with its conjugate, does indeed define a CKY 2-form, as can easily be checked. In the supersymmetric case one has
\be
D_{(\a i} K_{\b\c)}=0;\qquad \bar D_{\adt}^i K_{\b\c}=0\ .
\la{4.9}
\ee

This form is the first in a sequence of tensors of the form $K^{2+k,k}, k\in\bbN$. We claim that these tensors correspond to the CKYTs $A_{1,k+1}$. Clearly $K^{k+2,k}$ corresponds to the tensor product of a $k$-fold symmetric, traceless $D=4$ Lorentz tensor with a (complex) 
self-dual 2-form represented by the extra pair of undotted indices. Imposing total symmetry over the $(k+2)$ undotted indices gives an irreducible representation, which, when combined with its complex conjugate is indeed $A_{1,q+1}$. Equivalently, $A_{1,q} \sim K^{q+1,q-1}+ c.c$.

In the supersymmetric case each leading term $K^{2+k,k}$ descends to spinorial objects $\L^{1+k,k}$ and $\bar\P^{2+k,k-1}$ on applying $D$ and $\bar D$ respectively:
\begin{align}
D_{\a i} K_{\b_1\ldots \b_{2+k},\bdt_1\ldots\bdt_k}&=\ve_{\a(\b_1}\L_{\b_2\ldots\b_{2+k}) i,\bdt_1\ldots\bdt_k} \ , \nn\w1
\bar D_{\adt}^i K_{\b_1\ldots \b_{2+k},\bdt_1\ldots\bdt_k}&=\ve_{\adt(\bdt_1}\bar\P^i_{\b_1\ldots\b_{2+k}, \bdt_2\ldots \bdt_k)}
\label{4.10}
\end{align} Further descendants are obtained by repeated differentiation with the odd derivatives, leading to a picture similar to that for SCKTs given in \cite{Howe:2015bdd}. 
 In general we can arrange the $\th$-components of a general $D=4$ SCKYT in an array with each vertex labelled by a pair of integers, $(p,q)\leq (2+k,k)$, connected by arrows representing the action of $\bar D$ or $D$ acting respectively to the left or right down the diagram. The vertex $(p,q)$ therefore represents a tensor with $p\  (q)$ symmetrised undotted (dotted) spinor indices, and $(3-p)\ ((1-q))$ antisymmetrised lower (upper) internal indices. For example, for $k=1$, we have the diagram:

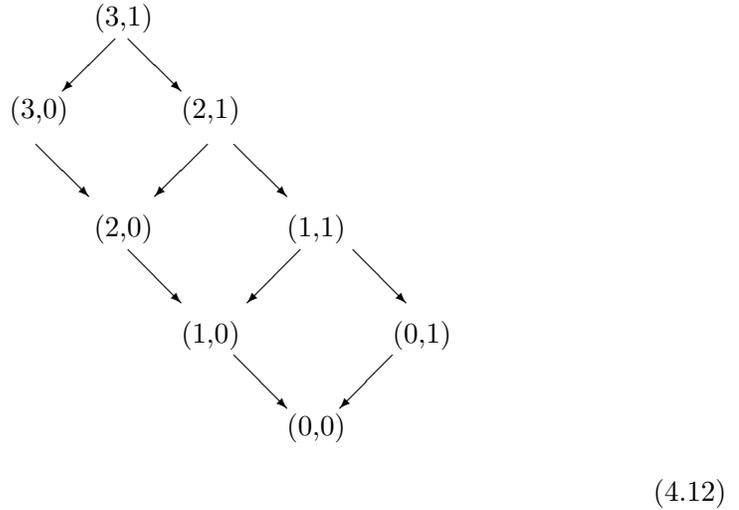
\be
\begin{picture}(400,200)
\put(202,180){(3,1)}
\put(210,175){\vector(-1,-1){20}}\put(215,175){\vector(1,-1){20}}
\put(170,145){(3,0)}\put(235,145){(2,1)}
\put(180,135){\vector(1,-1){20}}\put(245,135){\vector(-1,-1){20}}\put(255,135){\vector(1,-1){20}}
\put(202,100){(2,0)}\put(275,100){(1,1)}
\put(280,95){\vector(-1,-1){20}}\put(300,95){\vector(1,-1){20}}
\put(215,95){\vector(1,-1){20}}
\put(235,60){(1,0)}\put(315,60){(0,1)}
\put(255,55){\vector(1,-1){20}}
\put(315,55){\vector(-1,-1){20}}
\put(275,25){(0,0)}
\end{picture}
\la{4.13.b}
\ee

 \black

It is easy to compute dimensionalities in the spinor formalism. We find
\be
d_{1,q}=2q(q+2)\ ,
\la{4.11}
\ee
while the dimensionality of $D A_{1,q}$ is given by twice the dimensionality of $\L^{q,q-1}+\bar\P^{q+1,q-2}$. This is given by $4(q^2+q-1)$. The factor of $4$ is $d_s$, disregarding the internal symmetry algebra for the moment, so that this number is to be compared with the right-hand side of \eq{4.6}.	It is straightforward to verify that the dimensionalities computed in the general formalism and in the spinor formalism match up, as they should.

The next sequence up consists of tensors of the form $K^{4+k,k}, k\in\bbN$. The first term, $K^{4,0}$, together with its conjugate, is a tensor with the symmetries of the Weyl tensor. The object  $K^{4+k,k}$ together with its conjugate gives a tensor of the form $A_{ab,c_1\ldots c_k}$ which is symmetric on $ab$ and on the $c$s, but not symmetric on more than $k$ indices nor on $ab$ together with any $c$ index. It is also completely traceless. The Young tableau for such a representation is
\be
A_{1,1,q}\sim\ \ \ \ \ \overbrace{\young(\hfil\hfil\hfil\hfil\hfil,\hfil\hfil)}^{q}
\la{4.11}
\ee
where it is understood that all traces have been removed.

\subsection{$D=6$}


In six-dimensional spacetime the R-symmetry algebra for $N$-extended supersymmetry is $\gs\gp(N)$ (with $\gs\gp(1)\equiv \gs\gu(2)$), and the corresponding superconformal algebra is $\go\gs\gp(6,2)|N)$. The spin algebra is isomorphic to $\gs\gu^*(4)$, a non-compact version of $\gs\gu(4)$, and it is convenient  to represent tensors as well as spinors in terms of Young diagrams for this algebra. A rank-$n$ Killing tensor $K$ has the $\gs\gu^*(4)$ tableau
\be
K\sim \ \overbrace{\yng(7,7)}^{n}\ .
\la{4.12}
\ee
Applying a derivative $\del\sim\tiny\young(\hfil,\hfil)$ to $K$ the constraint is that the largest representation in $\del K$, \ie $\overbrace{\tiny\yng(8,8)}^{n+1}\ $,  should vanish.

A 2-form in $D=6$ has 15 components and is therefore represented by the tableau $\tiny\young(\hfil\hfil,\hfil,\hfil)$. The constraint obeyed by a CKY 2-form $B$ is that the largest representation in $\del B$ should vanish. This representation is 64-dimensional and has the $\gs\gu^*(4)$ tableau $\tiny\young(\hfil\hfil\hfil,\hfil\hfil,\hfil)$. This constraint is equivalent to the standard one \eq{2.1}. A self-dual 3-form $C^+$ has the tableau $\tiny\young(\hfil\hfil,\hfil\hfil,\hfil\hfil)$, while an anti-self dual 3-form $C^-$ has the tableau $\tiny\young(\hfil\hfil)$. In both cases the CKY constraint is that the largest representation in $\del C^{\pm}$ should vanish. Thus in both cases $\del C$ is a 2-form. The components of CKY 2- and 3-forms are
\begin{align}
(2)\rightarrow (3)&+(1)\rightarrow (2)\ ,\nonumber\\
(3)\rightarrow(4)&+(2)\rightarrow (3) \ .
\la{4.12.1}
\end{align}
In $D=6$, these give  respectively 56 and 70-dimensional representations of $\go(8)$, but the latter is reducible into self-dual and anti-self-dual 35-dimensional representations. In $D=6$ a CKY 3-form therefore splits into two: $(3)^{\pm}\rightarrow (4)\cong (2)\rightarrow (3)^{\mp}$.

In the supersymmetric case an $n$th rank SCKT is represented by the same Young tableau but with the constraint that applying an odd derivative $D_{\a i},i=1,\ldots 2N,$ to it  one finds
\be
D K\sim \ \overbrace{\young(\hfil\hfil\hfil\hfil\hfil\hfil\hfil,\hfil\hfil\hfil\hfil\hfil\hfil\hfil,\cdot)}^{n}\ ,
\la{4.13}
\ee
where $D$ is represented by the dotted box and where the other possible representation with an extra box on the first row is constrained to vanish. A putative superconformal SCKY 2-form $B$ satisfies the following constraint
\be
D B\sim \young(\cdot) \xz \young(\hfil\hfil,\hfil,\hfil)=\young(\hfil\hfil,\hfil,\hfil,\cdot)= \young(\cdot)\ ,
\la{4.14}
\ee
where, as in \eq{4.12}, the dot in a box also represents the fundamental representation of $\gs\gp(N)$. This constraint complies with the general definition of \eq{4.2}. However, it is too strong, because at the next level one finds the spacetime derivative of $B$ is a one-form together with an anti-self dual three-form, and this does not define a conformal representation. One might wonder whether one could relax the basic constraint \eq{4.2}, but it is easy to check that this would be too weak because then the large 64-dim representation in $\del B$ does not vanish.  In the case of 3-forms, only the self-dual case is compatible with supersymmetry. The constraint is
\be
D C^+\sim \young(\cdot) \xz \young(\hfil\hfil,\hfil\hfil,\hfil\hfil)= \young(\hfil\hfil,\hfil\hfil,\hfil\hfil,\cdot)\ ,
\la{4.15}
\ee
or
\be
D_{\a i} C^+_{abc}=(\c_{abc})_{\a\b} \chi^{\b}_i\ .
\la{4.16}
\ee
This last equation makes it clear that the anti-self-dual 3-form is incompatible  with the constraint \eq{4.2}.

So in $D=6$ one can only have self-dual 3-forms as SCKY forms. From these we can build tensors $A_{2,q}^+$ by taking the highest weight representation in the product of $C^+$ with a $(q-1)$th-rank SCKT. The $\gs\gu^*(4)$ tableau for the leading component is
\be
A_{2,q}^+\sim \overbrace{\young(\hfil\hfil\hfil\hfil\hfil\hfil\hfil\hfil,\hfil\hfil\hfil\hfil\hfil\hfil\hfil\hfil,\hfil\hfil)}^{q+1}
\la{4.17}
\ee
It obeys the constraint
\be
DA_{2,q}^+\sim \overbrace{\young(\hfil\hfil\hfil\hfil\hfil\hfil\hfil\hfil,\hfil\hfil\hfil\hfil\hfil\hfil\hfil\hfil,\hfil\hfil\cdot)}^{q+1}\ +\ 
\overbrace{\young(\hfil\hfil\hfil\hfil\hfil\hfil\hfil\hfil,\hfil\hfil\hfil\hfil\hfil\hfil\hfil\hfil,\hfil\hfil,\cdot)}^{q+1}\ .
\la{4.18}
\ee 
One can also have higher-rank SCKYTs by incorporating further pairs of 3-box columns into the Young tableau. For example,
\be
A_{2,2,q}^+\sim \overbrace{\young(\hfil\hfil\hfil\hfil\hfil\hfil\hfil\hfil\hfil\hfil,\hfil\hfil\hfil\hfil\hfil\hfil\hfil\hfil\hfil\hfil,\hfil\hfil\hfil\hfil)}^{q+3}
\la{4.19}
\ee

In order to extend these results to super Young tableaux it will be convenient  to consider dual $su^*(4)$ tableaux with a single box corresponding to an upper index. Thus $C^+\sim \tiny\young(\hfil\hfil)$. In indices,
\be
D_{\a i} C^{\b\c}=\d_\a{}^{(\b} \L^{\c)}_i\qquad {\rm where}\qquad \L^\a_i=\frac{2}{5} D_{\b i} C^{\b\a}\ .
\la{4.20}
\ee
The dual Young tableau for $A^+_{2,q}$ is then
\be
A_{2,q}^+\sim \overbrace{\young(\hfil\hfil\hfil\hfil\hfil\hfil\hfil\hfil,\hfil\hfil\hfil\hfil\hfil\hfil)}^{q+1}\ ,
\la{4.20.1}
\ee
while the constraint becomes
\be
D A_{2,q}^+\sim \overbrace{\young(\hfil\hfil\hfil\hfil\hfil\hfil\hfil\hfil\cdot,\hfil\hfil\hfil\hfil\hfil\hfil,\cdot,\cdot)}^{q+2}+ 
\overbrace{\young(\hfil\hfil\hfil\hfil\hfil\hfil\hfil\hfil,\hfil\hfil\hfil\hfil\hfil\hfil\cdot,\cdot,\cdot)}^{q+1}\ ,
\la{4.20.2}
\ee
where now $D$ is represented by a 3-box vertical tableau with dots.
\subsection{$D=5$}

In $D=5$ it is only for $N=1$ that there is a superconformal algebra, the exceptional superalgebra $\gf(4)$. The spin group is $\gs\gp(2)$, while the internal R-symmetry algegra is $\gs\gp(1)\cong \gs\gu(2)$. The spinor coordinates for conventional superspace are $\th^{\a i},\ \a=1,\ldots 4,\  i=1,2$. They are symplectic Majorana and have 8 real components. Since a 3-form is related to a 2-form by duality, we need only consider CKY 2-forms and CKYTs of type $A_{1,q}$. 

In $D=5$ it is again convenient to use spinor notation. An $q$th rank SCKT $K$  is represented by an $\gs\gp(2)$ Young tableau
\be
K\sim \ \overbrace{\young(\hfil\hfil\hfil\hfil\hfil\hfil\hfil,\hfil\hfil\hfil\hfil\hfil\hfil\hfil)}^{q}\ ,
\la{4.21}
\ee
where it is assumed, here and throughout this subsection, that all of the traces with respect to the $\gs\gp(2)$ symplectic form have been removed from all of the tableaux. The superconformal constraint is
\be
D K\sim \ \overbrace{\young(\hfil\hfil\hfil\hfil\hfil\hfil\hfil,\hfil\hfil\hfil\hfil\hfil\hfil\ast)}^{q}\ ,
\la{4.22}
\ee
where the asterisk denotes that the box containing it is {\sl removed} from the $\gs\gp(2)$ diagram, but carries an internal $\gs\gp(1)$ fundamental representation index coming from $D_{\a i}$. 

For a SCKY 2-form $B$ we have
\be
B\sim \young(\hfil\hfil)\ ,
\la{4.23}
\ee
while the constraint is
\be
D B\sim \young(\hfil\ast)\ ,
\la{4.24}
\ee
Combining these we get the diagram for a SCKYT $A_{1,q}$:
\be
A_{1,q}\sim \overbrace{\young(\hfil\hfil\hfil\hfil\hfil\hfil\hfil\hfil\hfil,\hfil\hfil\hfil\hfil\hfil\hfil\hfil)}^{q+2}\ .
\la{4.25}
\ee
The constraint is
\be
D A_{1,q}\sim \overbrace{\young(\hfil\hfil\hfil\hfil\hfil\hfil\hfil\hfil\hfil,\hfil\hfil\hfil\hfil\hfil\hfil\ast)}^{q+2}\  +\ 
\overbrace{\young(\hfil\hfil\hfil\hfil\hfil\hfil\hfil\hfil\ast,\hfil\hfil\hfil\hfil\hfil\hfil\hfil)}^{q+2}\ .
\la{4.26}
\ee
It is straightforward to check that these expressions are consistent with the general discussion given above.
\section{Analytic superspace}
In \cite{Howe:2015bdd} and \cite{Howe:2016iqw} we discussed SCKTs in analytic superspaces for $D=3,4,6$. {These are superspaces with fewer odd coordinates than the associated conventional (Minkowski) superspaces, such that superfields on these spaces correspond to fields on Minkowski superspace satisfying constraints with respect to the odd derivatives. These  superfields generalise the notion of chiral superfields, and typically also depend on additional internal even coordinates. Superspaces of this type were first introduced in the physics literature as harmonic \cite{Rosly,Galperin:1984av,Galperin:1984bu} or projective superspaces \cite{Karlhede:1984vr,Grundberg:1984xr,Lindstrom:2008gs}. More general treatments were later developed in \cite{Lukierski:1988vw,Hartwell:1994rp,Howe:1995md} where it was found convenient
to work in complexified superspaces defined as cosets of the superconformal groups with parabolic isotropy groups: these are flag supermanifolds \cite{Manin:1988ds,Harnad:1987xq,Harnad:1995zy}.  All the fields are taken to be holomorphic, and we shall usually work on some open subset in the spacetime sector as the cosets themselves are compact (in the even directions). The spaces we shall consider all contain standard complexified Minkowski space as a component of the purely even part, and indeed there is a formal resemblance to Minkowski spaces considered as cosets of the conformal groups.\footnote{One can consider these spaces as supersymmetric versions of twistor geometry, see, for example, \cite{Penrose:1986ca,Baston:1989vh,Ward:1990vs}.} They have additional even sectors, cosets of the R-symmetry groups, and reduced number of odd coordinates compared to Minkowski superspace. The analytic superspace formalism we shall use is one in which local coordinates are employed for all of the coordinates including the internal and odd ones. We shall be interested in those for which the reduction in the number of odd coordinates is maximal, and we shall also restrict our attention to the simpler cases of $N$ even, for $D=3,4$. For examples of this formalism applied to $N=4$ superconformal field theory see, for example, \cite{Heslop:2003xu}.} 

Analytic superspaces are thus particular coset superspaces of the complexified superconformal groups, $SL(4|N)$ for $D=4$, $OSp(8|N)$ for $D=6$ and $SpO(2|N)$ for $D=3$ the latter being isomorphic to $OSp(N|2)$ but written in the opposite order to indicate that $Sp(2)$ is the spacetime conformal group for $D=3$. These groups act naturally on the corresponding super-twistor spaces $\bbC^{4|N}$ and $\bbC^{8|2N}$ respectively. In $D=4$ we consider the Grassmanians which are spaces of $(2|M)$ planes in $\bbC^{4|N}$ where $N=2M$ or $N=2M+1$, depending on whether $N$ is even or odd (we exclude $N=1$, though). These coset spaces are in a sense the natural generalisations of complexified Minkowski spaces considered as cosets of the corresponding conformal groups. For example,  for $D=4$ Minkowski space is locally coordinatised by $x^{\a\a'}$ (we use primes instead of dots in this section), while the analytic superspaces have coordinates $X^{AA'}$, where $A=(\a, a)$, $a$ being an internal index which we take to be odd and $A'=(\a' ,a')$. 
\subsection{$D=4$}
We recall that in $D=4$ an $n$th-rank  CKT can be written in spinor notation as an object $K^{\a_1\ldots \a_n,\a'_1\ldots\a'_n}$ with $n$ primed and unprimed spinor indices which is totally symmetric on both sets of indices. In analytic superspace, we simply have to replace $(\a,\a')$ by $(A,A')$ and impose similar differential constraints to get a SCKT $K^{A_1\ldots A_n,A'_1\ldots A'_n}$. As have remarked previously, we can also consider objects $K^{m,n}$ with $m$ unprimed and $n$ primed indices, and these objects include the SCKYTs. The constraint obeyed by such tensors is

\begin{align}
\del_{AA'} K^{B_1\ldots B_m,B'_1\ldots B'_n}&= a_m( \d_A{}^{(B_1}\del_{CA'} K^{B_2\ldots B_m)C,B'_1\ldots B'_n})+
a'_n(\d_{A'}{}^{(B'_1}\del_{AC'} K^{B_1\ldots B_m,B'_2\ldots B'_n)C'})\nn\w1 
&+b_{m,n}\, \d_A{}^{(B_1}\d_{A'}{}^{(B'_1}\del_{CC'} K^{B_2\ldots B_m) C,B'_2\ldots B'_n)C'}\ ,
\label{6.1}
\end{align}
where
\be
a_m=\frac{1}{t_m} \qquad a'_n=\frac{1}{t'_n} \qquad b_{m,n}=-\frac{1}{t_m t'_n}\ ,
\la{6.2}
\ee
with 
\be
t_m=\frac{m-1+t}{n}\qquad t'_n=\frac{n-1+t'}{n}\ .
\la{6.3}
\ee 
Here $t=t'=2-M$ for $N$ even and $t=2-M,\ \ t'=2-(M+1)$ for $N=2M+1$. The numbers $t,t'$ are the super traces of the unit matrices $\d_A{}^B$ and $\d_{A'}{}^{B'}$ for the two halves of super-twistor space. 

The formula \eq{6.1} covers all of the SCKYTs in $D=4$ and indeed extends to include the supersymmetric extensions of SCKY spinors when $m+n$ is odd. Moreover, \eq{6.1} is easy to solve  as a power series in the local coordinates $X^{AA'}$. A simple example is given by the SCKY 2-form which represented by $K^{2,0}$ together with its conjugate $K^{0,2}$. The solution to \eq{6.1} for $K^{2,0}$ is
\be
K^{AB}=k^{AB} + X^{AA'} k_{A'}{}^B + X^{AA'} X^{BB'} k_{B'B} \ .
\la{6.4}
\ee
where the coefficients $k$ are constants. 

Denoting the coordinates for $D=4$ super-twistor space by $z^{\cA}=(z^A, z_{A'})$ we can assemble the above components into a symmetric second-rank tensor $K^{\cA\cB}$ in twistor space  by
\be
K^{\cA\cB}=\left(k^{AB}, k^A{}_{B'}, k_{A'}{}^B, k_{A'B'}\right)\  ,
\la{6.5}
\ee
where $k_{A'}{}^B=k^B{}_{A'}$. This exhibits the SCKY 2-form manifestly as a representation of the superconformal algebra.

The higher-rank SCKYTs, all of the form $A_{1,q}$ in $D=4$, are given by tensors $K^{m,n}$ obeying \eq{6.1} with $m=q+1, n=q-1$ together with their conjugates $K^{n,m}$.  The super-twistors corresponding to these objects have the form
\be
K^{\cA_1\ldots \cA_{m}}{}_{\cB_1\ldots \cB_{n}}=K^{\cA_1\ldots \cA_{q+1}}{}_{\cB_1\ldots \cB_{q-1}}\ ,
\la{6.6}
\ee
and are totally symmetric on both the upper and lower sets of indices. Again, the components of these objects are constant. In addition, in order to get the full tensor which will correspond to the real version in super Minkowski space it will be necessary to include the conjugate object, $K^{n,m}=K^{q-1,q+1}$. 

\subsection{$D=6$}


In $D=6$ analytic superspace is the space of isotropic (with respect to the standard metric) $(4|N)$ planes in $\bbC^{8|2N}$. Local coordinates are $X^{AB}=(x^{\a\b},\xi^{\a b},y^{ab})$, where $x^{\a\b}$ are the spacetime coordinates, $y^{ab}$ are internal even coordinates and $\xi^{\a b}$ are the odd coordinates; here greek indices run from 1 to 4 and latin indices from 1 to $N$. So this space has half the number of odd coordinates of super Minkowski space and has  additional even coordinates. The internal part of the space is the coset $U(N)\bsh Sp(N)$.

 In $D=6$ we can take over most of the results of the even case, as far as the representations are concerned, by simply interpreting the Young tableaux to be those of $\go\gs\gp(8|N)$. In particular, a $q$th rank SCKT is given by the diagram \eq{4.12} (with $n$ replaced by $q$). For the self-dual 3-form we need to use the dual tableau representation $\tiny\young(\hfil\hfil)$, so that the generalised products will be given by diagrams of the type \eq{4.20.1}.  The differential constraint satisfied by the super 3-form $C^{AB}$ in analytic superspace is simply
\be
\del_{AB} C^{CD}= -\frac{4}{t} \del_{[A}{}^{(C} \del_{B]E} C^{D)E}\ ,
\la{6.8}
\ee
where $t$ is the super-dimension of the tangent space in analytic superspace. 
 For SCKTs, the answer was given in \cite{Howe:2015bdd}; it is

\begin{align}
\del_{A_1 A_2} K^{B_1 B_2,C_1, C_2,\ldots} &=(a_n\, \d_{[A_1}{}^{[B_1} \del_{A_2] D} K^{B_2] D,C_1 C_2,\ldots} +
(n-1)\  {\rm terms})\nn\w1
&+ b_n\,(\d_{[A_1}{}^{[B_1}\d_{A_2]}{}^{B_2]}\del\cdot K^{C_1C_2,\ldots}+ {\rm cyclic})\nn\w1
&-\frac{6b_n}{n+1}(\sum \d_{[A_1}{}^{[B_1}\d_{A_2]}{}^{B_2} \del\cdot K^{C_1 C_2],D_1D_2,\ldots})\ ,
\la{5.5}
\end{align}
where in the second line the cyclic sum is over the $n$ pairs, and where the sum in the third line is over all distinct pairs of pairs, \ie $\half n(n-1)$ terms altogether. In the expression on the third line for each selected pair of pairs there is total graded antisymmetrisation. It can be checked that these terms are necessary to ensure that the (graded) symmetry structure of the tableau \eq{4.11} holds for the $b$ terms, while the $a$ terms take care of themselves. We have used the dot notation to denote the divergence with respect to a given pair of indices. The coefficients are given by
\begin{align}
a_n&=\frac{4}{t+n-3}\nn\w1
b_n&=\frac{-(n+1)}{(t+n-2)(t+n-3)}\ .
\la{5.6}
\end{align}

Rank $n$ SCKTs of the above type correspond to the Young tableaux $\overbrace{\tiny\yng(8,8)}^{n}\ $, each pair of indices on $K$ in the above formula \eq{5.5} corresponding to a given column. However, when we construct SCKYTs by taking the highest weight representations of products of SCKTs with several SCKY 3-forms, the resulting Young tableaux have the form 
\begin{align}
&\overbrace{\yng(10)}^m \nn\\[-4pt]
&\underbrace{\yng(8)}_{n}\ ,  \qquad\qquad m\geq n\ .
\la{5.7}
\end{align}
In fact we can be quite general here and allow $m, n$ to be arbitrary non-negative integers, thereby including SCKYT-spinors. In index notation, such a tableau translates to a tensor $K^{A_1,\ldots A_m, B_1\ldots B_n}$ which is symmetric on both sets of indices but not on any more than $m$, so that
\be
K^{(A_1\ldots A_m,B_1) B_2\ldots B_n}=0\ .
\la{5.8}
\ee
These symmetry properties are enough to specify the tableau \eq{5.7}. Note that the algebras here are $\gg\gl$ or $\gs\gl$ so that there are no metric traces that can be removed. The constraints obeyed by such a tensor (spinor) in order for it to be a superconformal one are:

\begin{align}
\del_{AB} K^{C_1\ldots C_m,D_1\ldots D_n}&=a_{m,n} \d_{[A}{}^{(C_1}\del_{B]E} K^{|E|C_2\ldots C_m) ,D_1\ldots D_n} \nn\w1&+
b_{m,n}\d_{[A}{}^{(C_1}\del_{B]E}K^{C_2\ldots C_m) (D_1,D_2\ldots D_n) E}\nn\w1
&+c_{m,n} \d_{[A}{}^{(D_1} \del_{B]E} K^{|C_1\ldots C_m|,D_2\ldots D_n) E} \nn\w1 &+e_{m,n}\d_{[A}{}^{(C_1}\d_{B]}{}^{(D_1} \del_{EF} K^{C_2 \ldots C_m) E, D_2\ldots D_n )F}\ ,
\la{5.9}
\end{align}
where in the last line the $C$s and $D$s are separately symmetrised. The coefficients are determined by taking traces and double traces and by requiring that the symmetry of \eq{5.8} is satisfied by the right-hand side terms. The result is:
\begin{align}
a_{m,n}&=-\frac{2m}{t+m-2}\nn\w1
b_{m,n}&=\frac{2mn}{(t+m-2)(t+n-3)}\nn\w1
c_{m,n}&=\frac{-2n}{t+n-3}\nn\w1
e_{m,n}&=-\frac{2mn}{(t+m-2)(t+n-3)}\ ,
\la{5.9}
\end{align}
where $t=4-N$ is the supertrace $\d_A{}^A$ (= half the super-dimension of super-twistor space). Note that the discussion applies equally well to the non-supersymmetric case where the algebra is $\gs\gl(4)$ in the complexified case.

The discussion of these tensors as representations of $\go\gs\gp(8|N)$ is straightforward. One takes the same Young diagrams used in either super-Minkowski space or in analytic superspace and reinterprets them as Young tableaux in $\go\gs\gp(8|N)$, with the additional requirement that these tensors be completely traceless with respect to the orthosymplectic metric. For example, in the purely even case a Killing tensor in spacetime {is given in spinor notation by \eq{4.12}, and this becomes the same diagram} in $\go(8)$ but with the traces removed. This discussion remains valid for the SCKYTs.

\subsection{$D=3$}

$D=3$ analytic superspace, for $N=2M$ even,  is the space of isotropic $(2|M)$-planes in $\bbC^{4|M}$, isotropic being with respect to the orthosymplectic metric on $\bbC^{4|2M}$ regarded as a supersymplectic two-form, because the spacetime part is the symplectic part in this case. The local coordinates are given by $X^{AB}$, where $A=(\a,a)$ with $\a=1,2,3$   while the internal index $a$ runs from 1 to $M$. The super-coordinates $X^{AB}$ in this case are graded symmetric. As a 2-form in $D=3$ is dual to a 1-form it follows that there are no independent CKY forms, while a 1-form is just a CKV. We include this case in order to generalise the formula given for {SCKTs in \cite{Howe:2015bdd}} to the case of SCKYT-spinors. Such an object is a graded-symmetric $m$th rank tensor $K^{A_1\ldots A_m}$ subject to the constraint
\be
\del_{A_1 A_2} K^{B_1\ldots B_m}=a_m\d_{(A_1}{}^{(B_1}\del_{A_2)C} K^{B_2\ldots B_m)C} + b_m \d_{(A_1}{}^{(B_1}\d_{A_2)}{}^{B_2}\del_{CD} K^{B_3\ldots B_m)CD}\ ,
\ee
where
\be
a_m=\frac{2m}{t+m}\qquad b_m=-\frac{m(m-1)}{(t+m)(t+m-1)}\ .
\ee
This agrees with the formula given for $n$th rank SCKTs in \cite{Howe:2015bdd} when $m=2n$.

\subsection{Decomposability}

It is well-known that there can be representations of super Lie algebras which contain sub-representations which cannot be removed because the subtraction process involves super-traces which can vanish  in some cases \cite{Bars:1982se,Leites:2002}. In \cite{Howe:2015bdd} we discussed examples of this in the context of superconformal Killing tensors (SCKTs), but clearly problems of this sort can also arise for SCKYTs and SCKYT-spinors. This is most simply discussed in super-twistor spaces where these objects are represented by tensors with constant components. The restrictions are very similar to those for SCKTs: in particular, problems only arise for values of $N$ which are less interesting from a physical point of view because there are no interacting superconformal field theories, with one exception.

For $D=3$ a SCKYT-spinor is given by a totally symmetric rank $m$ tensor on twistor space $\bbC^{4|N}$. In this case the invariant tensor is the supersymplectic 2-form and hence there are no subrepresentations to worry about.

For $D=6$ a SCKYT-spinor is given by a tensor in $\bbC^{8|2N}$ corresponding to a tableau of the type depicted in \eq{5.7}, but assumed to be traceless with respect to the orthosymplectic metric. Here there can be cases where subrepresentations cannot be removed. A simple example is given by the tableau with two rows and columns, which has the symmetries of the Riemann tensor, so that one would expect to form an irreducible tensor with the symmetries of the Weyl tensor by removing the traces. However, this is not possible for $N=3$, and in fact problems of this type do not occur for $N<3$, i.e. they are absent in the most interesting cases from a field theory point of view.

For $D=4$ problems can start at $N=4$, but these only involve single traces. A fuller discussion of this topic can be found in {\cite{Howe:2015bdd}}.

\section{Conclusion}

In this paper we have studied some aspects of conformal Killing-Yano tensors from an algebraic point of view and shown how these tensors  naturally lead to invariants of classical spinning particles. In the quantum case such classical invariants define the leading terms of higher symmetries of the differential operators that arise as the quantised first-class constraints. For $N_0=1$ worldline supersymmetry the relevant operator is the Dirac operator and our work here systematises that of some earlier investigations. The discussion in the $N_0=2$ case is more complicated, but our definition of general CKYTs clarifies their use in the construction of invariants. The alternative definition proposed earlier in \cite{Kress} has also been used in studying invariants of generalised Maxwell equations, which result from quantisation, particularly in four dimensions. In the case of the Dirac operator we also briefly discussed how general higher-order symmetries can be constructed from this point of view. However, further work needs to be done to obtain a complete understanding of this problem since the different types of spinor that can arise in different dimensions of spacetime is not fully taken into account in the basic particle model we have investigated here.

In the second part of the paper we extended the analysis of CKYTs to the supersymmetric case, i.e. to tensors of this type living in various superspaces.  The results here extend those that we discussed in earlier work, where we focused on superconformal Killing tensors. One question not addressed here is for which systems do these tensors arise as symmetries of differential operators. For SCKTs we know that these arise as symmetries of minimal superconformal models and that these can be interpreted in analytic superspaces as symmetries of super-Laplacian operators.\footnote{For some related work on this topic see \cite{Siegel:1999ew,Hatsuda:2008pm,Ju:2013mkc}.} The generalised SCKYTs arise as symmetries of non-minimal models which can be described by super Dirac equations in analytic superspaces, although there is no super analogue of the generalised Maxwell equation symmetries as these equations arise as components of either super Laplacians {\cite{Howe:2016iqw}} or of super Dirac operators {\cite{HL}}.

In the paper we have focused on flat spaces and superspaces but we have restricted our study to those systems which are compatible with non-trivial backgrounds spacetimes. For ordinary spacetimes this restriction is to particles with $N_0<3$ worldline supersymmetry. In the spacetime supersymmetric case, it is well-known that superconformal groups exist only for $D=3,4,6$ and one example in $D=5$, and in these cases SCKYT-spinors can always be defined.  However, one can only have interacting conformal supergravity backgrounds for limited values of $N$, the number of supersymmetries, and we have shown that for these cases the problem of indecomposable representations does not arise (except mildly for $D=4, N=4$).

In the purely bosonic case symmetries of the Laplacians  give rise to algebraic structures \cite{Eastwood:2002su} which play a role in higher spin theories
 \cite{Fronsdal:1978rb,Fradkin:1987ah,Vasiliev:1999ba} via the AdS/CFT correspondence {\cite{Mikhailov:2002bp}. It would therefore be of interest to understand whether extended algebras of this type could arise in the case of symmetries of Dirac and other operators, but a fuller discussion would require a complete understanding of the higher symmetries of these operators which we have not attempted to give here.
\bigskip

\noindent
{\bf Acknowledgement}:  U.L. gratefully acknowledges
the hospitality of the theory group at Imperial College, London, as well as support from the EPSRC
programme grant  ``New Geometric Structures from String Theory'' EP/K034456/1.

\black


\end{document}